\documentclass[12pt]{iopart}
\usepackage{iopams,amssymb,graphicx}

\newcommand{\Res}[3]{\mbox{Res}_{#1}^{#2}#3}
\newcommand{\Ind}[3]{\mbox{Ind}_{#1}^{#2}#3}
\newcommand{\Hom}{\mbox{Hom}}
\newcommand{\doublerightarrow}{\; -\!\!\! -\!\!\!\!\!\! \gg \;}
\newcommand{\bigplus}{\mbox{\LARGE $+$}}

\newcommand{\TL}{Temperley-Lieb}
\newcommand{\hash}{\#}
\newcommand{\BMW}{Murakami-Birman-Wenzl}
\newcommand{\BBase}{B^J}
\newcommand{\bra}[1]{| #1 \rangle}
\newcommand{\ket}[1]{ \langle #1 |}
\newcommand{\bigra}{{\mathcal G}}
\newcommand{\Simp}{{\mathcal S}}
\newcommand{\inT}{T_{n,\Z}}
\newcommand{\opA}{{\mathcal A}}
\newcommand{\opB}{{\mathcal B}}
\newcommand{\Gram}{Gram}
\newcommand{\sT}{T'}
\newcommand{\inc}{{\mathcal I}}
\newcommand{\Nsites}{n}

\newcommand{\C}{\mathbb{C}}
\newcommand{\N}{\mathbb{N}}
\newcommand{\Z}{\mathbb{Z}}
\newcommand{\R}{\mathbb{R}}
\renewcommand{\paragraph}{\refstepcounter{paragraph}
\noindent {\bf (\theparagraph)}\indent}
\renewcommand{\theparagraph}{\arabic{section}.\arabic{paragraph}}

\newtheorem{theo}{Theorem}
\newtheorem{pr}{Proposition}

\begin{document}
\jl{1}

\title[The Bubble Algebra]{The Bubble Algebra:\\
 Structure of a Two-Colour Temperley-Lieb Algebra}

\author{Uwe Grimm\dag\ and Paul P Martin\ddag}

\address{\dag\ Applied Mathematics Department, The Open University,
Walton Hall, Milton Keynes MK7 6AA, UK}

\address{\ddag\ Mathematics Department, City University, 
Northampton Square, London EC1V 0HB, UK}

\setcounter{footnote}{2}

\begin{abstract}
We define new diagram algebras providing a sequence of multiparameter
generalisations of the Temperley-Lieb algebra, suitable for the
modelling of dilute lattice systems of two-dimensional Statistical
Mechanics.  These algebras give a rigorous foundation to the various
`multi-colour algebras' of Grimm, Pearce and others.  We determine
the generic representation theory of the simplest of these algebras,
and locate the nongeneric cases (at roots of unity of the
corresponding parameters).  We show by this example how the method
used (Martin's general procedure for diagram algebras) may be applied
to a wide variety of such algebras occurring in Statistical Mechanics.

We demonstrate how these algebras may be used to solve the
Yang-Baxter equations.
\end{abstract}


\section{Introduction}
\setcounter{paragraph}{0}

Some time ago, motivated by the study of dilute lattice models
\cite{R92,WNS92}, Grimm and Pearce \cite{GP93} introduced
generalisations of certain diagram algebras (algebras with a
diagrammatic formulation \cite{M91}), such as \TL\ \cite{TL71} and
\BMW\ \cite{M87,BW89} algebras.  These algebras are important in the
theory of solvable lattice models of two-dimensional Statistical
Mechanics \cite{Baxter} and are related to link and knot invariants
\cite{WDA89}. The generalisation was conceived on the diagram level by
introducing diagrams with lines in a number of colours.  Each algebra
was then described by generators and relations dictated, or at least
suggested, by the requirement of solving the Yang-Baxter equations.
However the diagrammatic (which is to say, topological) underpinning
was not precisely formalised.

The classes of solvable lattice models called dilute lattice models
\cite{R92,WNS92,ZPG95} whose discovery motivated this generalisation
are closely linked to models of dilute loops on a lattice
\cite{BNW89,WN93}.  These models attract particular interest because
they contain a solvable `companion' of the two-dimensional Ising model
in a magnetic field \cite{WNS92,GN96,GN97} --- one of the famous
unsolved problems in Statistical Mechanics. The idea here is to
consider two colours, and to regard the second colour merely as a
dilution of the first.

In the two colour case the requirement of solving the Yang-Baxter
equations is fully satisfied by the design of the relations.  The
Yang-Baxter equations are sufficient to guarantee solvability in the
sense of commuting transfer matrices \cite{Baxter,WDA89}.  Thus
representations of such algebras give rise to solvable dilute and
two-colour lattice models.  (More precisely, one has relations for a
{\em tower}\/ of algebras, and the representation must be defined for
the whole tower.)  Various explicit representations and associated
models are considered in \cite{GP93,G94a,G94b,GW95a,GW95b,G96}.

The representations found included previously known lattice models
\cite{G94a,G94b}, but also gave rise to new series of solvable lattice
models \cite{GW95a,GW95b,G96}.  However, little else was discovered
about these algebras and their structures.  We have generators and
relations, and enough representations to show that these relations do
not imply a trivial algebra, but no knowledge of dimensions or even
finiteness, and no analysis of irreducible representations.  To this
extent the representations which {\em were}\/ found were a matter of
luck, and there was no way to tell if the relations could engender
other important but undiscovered models.  This may be contrasted with
our quite complete knowledge of the representation theory of the
Temperley-Lieb algebra itself, which is strikingly rich and
beautiful, and important in several areas of mathematics and physics
\cite{DonkinX,KS92a,KS92b,M91,JonesX,Khovanov}.

In this paper we define a new algebra --- the {\em bubble}\/ algebra.
We define this algebra entirely diagrammatically, such that it is
amenable to the general method of \cite[\S9.5]{M91},\cite{MS94b}.  We
then show that this gives a properly constituted diagrammatic
realisation of the Grimm-Pearce multi-colour \TL\ algebra (i.e.\ it
solves the Yang-Baxter equations).  We hence use the general method
to determine the generic representation theory of these algebras
completely.  We set up the machinery to investigate their exceptional
representation theory (analogous to that of ordinary Hecke algebras at
$q$ a root of unity).  We show how irreducible representations may be
associated to physical observables in the corresponding lattice
models.  We conclude with a discussion of the implications of our
results for Bethe ansatz on models derived using this algebra.  We
mainly discuss the case of two colours, as the further generalisation
to more colours is straightforward.  (The case of one colour is the
original \TL\ algebra.)

Generalisations of the \TL\ algebra are two-a-penny
\cite{MS94a,MS94b, HCLee92,RuiXi,RGreen98}, however there are now a
number of reasons for looking at the algebras introduced in
\cite{GP93} again.  Firstly, the diagram form of the \TL\ algebra is a
deep and powerful property (cf. \cite{KS92a,KS92b,M91}), and our new
realisation provides a natural generalisation on the diagram level.
Secondly, they provide solutions of the Yang-Baxter equation as we
have said.  They are similar in some ways to the blob algebra, which
has recently been shown \cite{DM02} to be useful in solving the
reflection equation \cite{Sklyanin}.  We also expect them to be of use
in constructing integrable boundary conditions for certain solvable
lattice models, including `conformal twisted' boundary conditions
\cite{GS93,BP01,PZ01,G02,G03}, and thus to be of relevance to boundary
conformal field theory.  Thirdly, we show that they are part of a
class of algebras amenable to the methods of \cite{MS94b}, so that we
may now analyse them quite efficiently (and hence provide a uniform
theory of such algebras).  This analysis suggests (see later) that
they may be relevant for Statistical Mechanics on ladders (cf. recent
works \cite{WangSchlottmann,Wang,Tonel}), and indeed, just recently,
this was shown to be the case \cite{GB03}.  They also look like they
should be relevant for circuit design and even transport network
design (although we know of no example of their use in these areas!)
as we will see.  There are also similarities with \BMW\ algebras
\cite{M87,BW89} and Fuss-Catalan algebras \cite{DF98}, both of which
have been used to construct integrable systems, to the extent that the
same methods are applicable there.  Finally, they have a number of
features of technical interest in representation theory (we largely
postpone comment on these to a separate paper, but see section~6 for a
brief discussion).

We start with some definitions. 

\section{Diagram algebras}

Our new algebra is a diagram algebra --- an algebra with a
diagrammatic formulation akin to the Temperley-Lieb algebra
\cite{TL71,M91}.  It will be convenient to recall this familiar
example in a suitable formalism, and then generalise to our case.

\paragraph{}
Fix a rectangular subset of $\R^2$ such that there is an edge with a
North pointing normal (e.g. $[0,1]\times[0,1]$).  Label each edge by
the direction (NSEW) of its normal.  Consider the set of partitions of
this rectangle by finitely many continuous non-crossing lines (walls)
with no wall touching the E or W edge.  We define an equivalence
relation on this set by equivalencing two such partitions if they
differ only by a continuous edge preserving deformation of the
rectangle.  We call (representatives of) equivalence classes {\em
diagrams}.

We say we can {\em compose}\/ two such partitions, $a$ over $b$, if
there lie in their equivalence classes two diagrams such that when $a$
is juxtaposed with $b$ from above, the southern endpoints of lines in
$a$ coincide with the northern endpoints of lines in $b$ (NB, this
requires only that the {\em number}\/ of lines matches up).  Each
point of coincidence may then be regarded as an interior point of a
continuous line passing though the juxtaposition $a|b$.  The composite
$ab$ is the new partition of the combined region which results from
this.

\paragraph{}
Consider the subset of diagrams where there are precisely $n$
endpoints on each northern and each southern edge.  For $q$ an
invertible indeterminate, consider the $\Z[q,q^{-1}]$-linear
extension of this set.  Let $\inT$ denote the quotient of {\em this}\/
set by the relation which equivalences any diagram with a closed
(interior) loop to $\delta$ times the same diagram without, where
$\delta = q+q^{-1}$.  Note that $\inT$ has basis the set of diagrams
where there are precisely $n$ endpoints on each northern and each
southern edge, {\em and no interior loops}.  Note that the composition
of diagrams passes to a well defined composition on this set, making
it a $\Z[q,q^{-1}]$-algebra.

Fix $K$ a field which is a $\Z[q,q^{-1}]$-algebra (for example, the
complex numbers, with $q$ acting as some specified nonzero complex
number).  The {\em \TL\ algebra}\/ $T_n(q)$ is the $K$-algebra $K
\otimes_{\Z[q,q^{-1}]} \inT$.

\paragraph{}
A line in a diagram with one endpoint in the Northern (N) edge and one
in the S edge is called a {\em propagating}\/ line.  The identity
element of $T_n(q)$ is the unique diagram all of whose lines are
propagating.

There are a number of ways of embedding $T_{n-1}$ as a subalgebra in
$T_n$. We will call that embedding which maps $a \in T_{n-1}$ to the
same diagram, but with one extra propagating line on the right, the
{\em natural}\/ embedding.

\medskip

For brevity we will assume familiarity with the usual presentation of
$T_n$ by generators $\{U_1,U_2,..,U_{n-1} \}$ and relations, and the
correspondence with the diagram version (see for example \cite{M90}).

\medskip

The topological/diagram realisation of the \TL\ algebra is enormously
useful \cite{M91,RGreen98} and deep \cite{KS92a,KS92b,Khovanov}.  We
require a similarly clearcut and intuitive construction, let us call
it a {\em model}, for the algebra introduced in \cite{GP93}.  Here we
will concentrate mainly on the model for two colours.  The
generalisation to arbitrarily many colours will be obvious.  A
generalisation to the \BMW\ version is also possible.

Before we introduce the model note that the \TL\ diagrams described
above may be regarded as partitionings of the set of endpoints into
pairs. The non-crossing rule means that they are a proper subset of
the set $\BBase_n$ of all such pair partitionings in general.  The
full set $\BBase_n$ is a basis for the Brauer algebra $J_n$ \cite{B37}
(whose composition rule need not concern us here).

\paragraph{}
Now consider the set each element of which consists of two independent
(but simultaneous) partitionings of a rectangle as above (one, say,
with red lines, one with blue).  Here independence means that walls of
different colours may cross, but we will {\em exclude}\/ elements in
which such crossings occur on the frame of the rectangle.  We define
an equivalence essentially as before, so for example (locally)
\[
\includegraphics{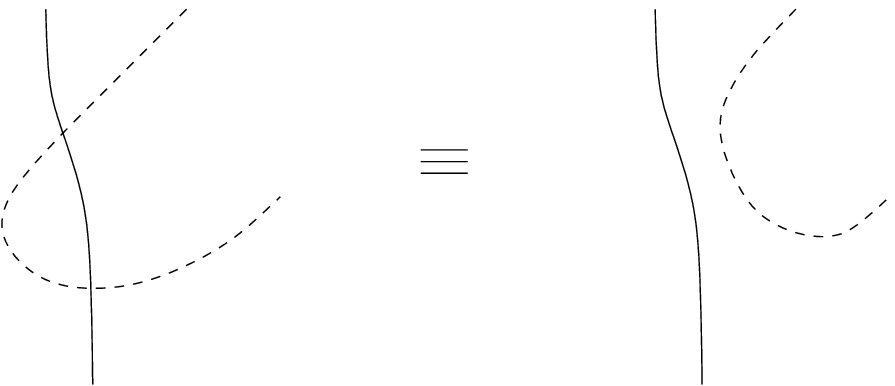}
\]
but (because of the exclusion) not on the frame:
\[
\includegraphics{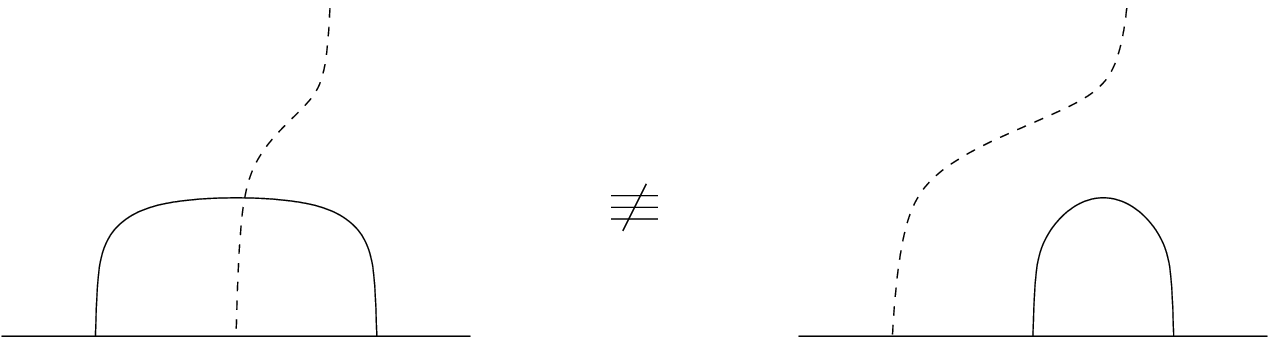}
\]
This is as if we have two parallel but independent deformable
rectangles (one for each colour), but they share the frame.  Another
way to think of this is as lines embedded not just in a rectangle, but
in bubble wrap (bubble wrap is made from two sheets of polythene
welded together along certain lines to trap bubbles).  Red lines are
allowed on the welds and the back sheet, blue lines are allowed on the
welds and the front sheet:
\[
\fl\includegraphics{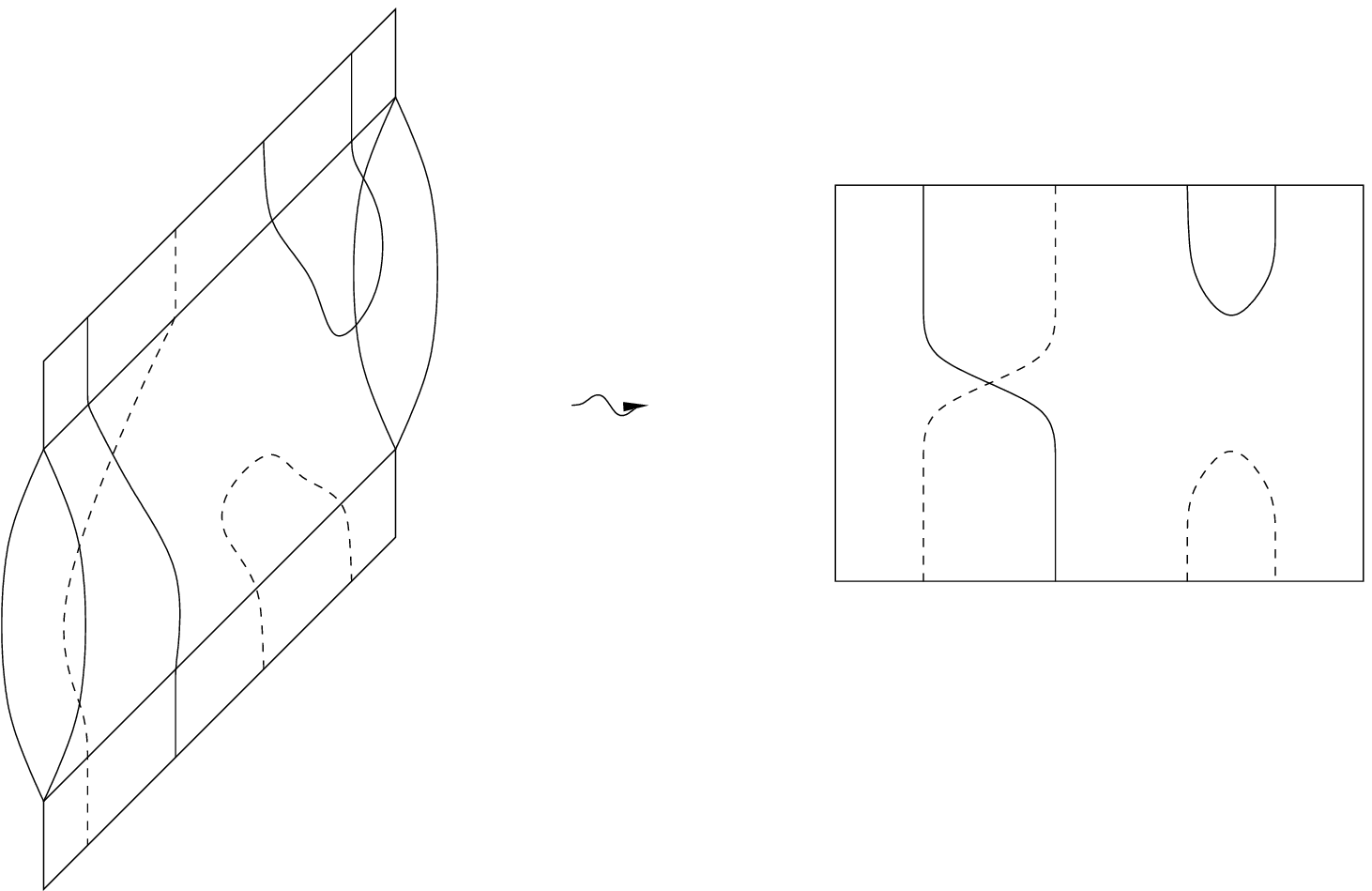}
\]
In this realisation lines on the same sheet (or on the weld) are not
allowed to touch, but otherwise may be deformed isotopically as
before.  Accordingly, we call the deformation equivalence `bubble
isotopy'.

Again we define composition whenever the number of endpoints
(irrespective of colour) matches up.  We do this as follows.  We call
the match up {\em precise}\/ if the colours match up precisely (i.e.\
we can identify the touching edges and have a properly formed
two-colour partition).  The composite is 0 unless the colours match
up precisely.  If they {\em do}\/ match up the composite {\em is}\/
that two-colour partition.

Consider the subset of double partitionings in which the total number
of endpoints (red {\em and}\/ blue) on the northern edge is $n$, and
similarly on the southern edge.  The {\em bubble algebra}\/ $\inT^2$
(so named to emphasise the topological diagram underpinning) is the
$\Z[q_r,q_r^{-1},q_b,q_b^{-1}]$-linear extension of this set and
composition, with internal closed loop replacements (as in $T_n(q)$).
Thus $\inT^2$ has a basis, $B_n$ say, of two-colour partitions (up to
bubble isotopy) with no internal loops.  The loop replacement scalar
$\delta$ here depends on the colour: $\delta_r = q_{r}+q^{-1}_{r}$ and
$\delta_b = q_{b}+q^{-1}_{b}$.  Fix a field $K$ which is a
$\Z[q_r,q_r^{-1},q_b,q_b^{-1}]$-algebra as before (e.g.\ the complex
numbers with $q_r,q_b$ specified complex numbers).  Denote the
$K$-algebra $K \otimes_{\Z[q_r,q_r^{-1},q_b,q_b^{-1}]} \inT^2$ by
$T^2_n = T^2_n(q_{r} , q_{b})$.

The obvious generalisations $T^N_n$ ($N=1,2,..$) include $T^1_n=T_n$.

\paragraph{}
Let $\hash_{r}(d)$ denote the number of {\bf r}ed propagating lines in
diagram $d$ (and similarly for {\bf b}lue).  Extend this to apply to
any non-zero scalar multiple of $d$.  It will be evident that
composing with any second diagram $d'$ such that $dd' \neq 0$ we have
\begin{equation}\label{prop_no}
 \hash_{r}(dd') \leq \hash_{r}(d) 
\end{equation}
and similarly for blue.  Write $B_n(i,j)$ for the subset of $B_n$ with
$\hash_{r}(d)=i$, $\hash_{b}(d)=j$, and define
\[
B_n(i)= \cup_j B_n(i-j,j)  
\]
\[
B_n[i]= \cup_{j \leq i} B_n(j)  
\]
so $B_{n}(i)$ and $B_{n}[i]$ consist of those diagrams in $B_{n}$ with
exactly $i$ and at most $i$ propagating lines, respectively.

We say that two lines are {\em strictly}\/ non-crossing when they are
non-crossing even when projected into a single plane (so as to
recover Grimm and Pearce's original diagrams).  Write $B'_n(i,j)$ for
the subset of $B_n(i,j)$ with lines all strictly non-crossing, and
define $B'_n(i)$ similarly.

For example $B'_n(n)$ is the set of diagrams with all lines
propagating and strictly non-crossing.  It will be evident that
$|B'_n(n)|=2^n$, and that
\[ 1 = \sum_{d \in B'_n( n)}   d
\]
is an orthogonal idempotent decomposition of the identity element of
$T^2_n$.

If $d \in B_{n-1}$, let $\inc_r(d) , \inc_b(d) \in B_n$ denote the
same diagram except with one extra non-crossing propagating red
(resp. blue) line to the right of all other lines.  Thus $\inc_r$ and
$\inc_b$ are injective maps on bases, which extend to injective maps
from $T^2_{n-1}$ to $T^2_n$. Note that these maps do not preserve the
identity element.  There is, however, an inclusion
\[
\inc : T_{n-1} \hookrightarrow T_n
\] 
given by  
$ d \mapsto \inc_r (d) + \inc_b (d) $ which we will call the `natural'
inclusion by analogy with the $T_n$ case.

\paragraph{} 
The basis $B_n$ may be constructed systematically for each $n$ from
that for $n-1$ using some simple combinatorial devices which we will
describe shortly.

Examples: The basis $B_1$ of $T^2_1$ consists of the following
diagrams
\[
\includegraphics{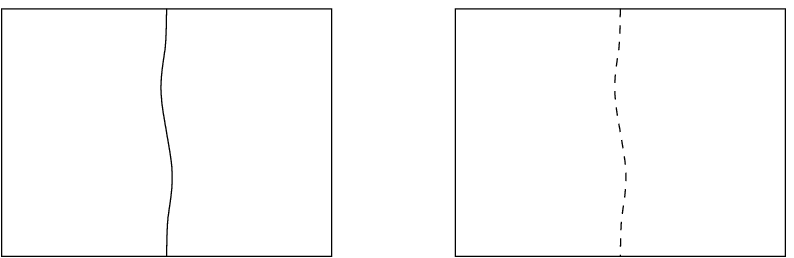}
\]

The basis $B_2$ of $T^2_2$ consists of the following diagrams
\[
\includegraphics{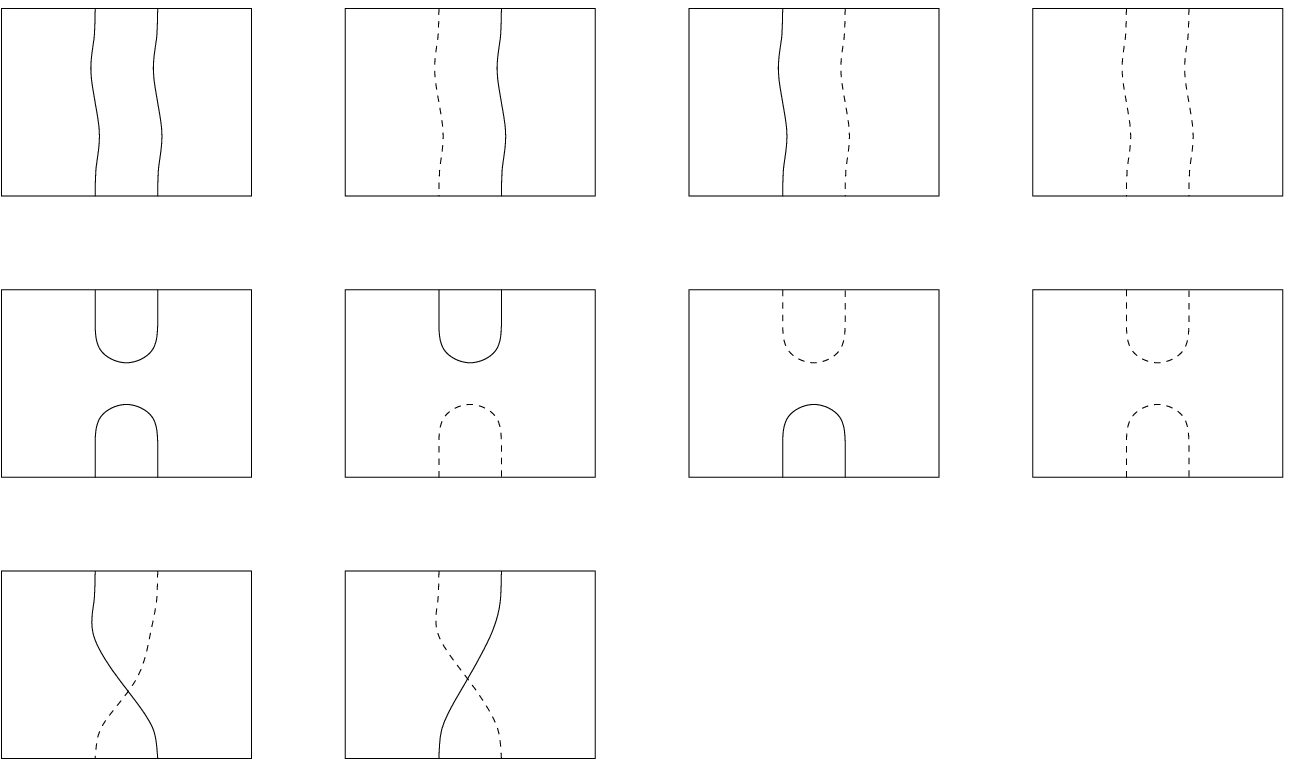}
\]
In particular $B_2(0,0)$ consists of the diagrams in the middle row;
$B_2(2,0) $ consists only of the leftmost diagram in the top row; and
$ B_2(0,2)$ the rightmost.  The remaining diagrams are in $B_2(1,1)$,
thus $B_2 = B_2(0,0) \cup B_2(2,0) \cup B_2(0,2) \cup B_2(1,1)$.

Let us write $U^r_1$ for the rightmost diagram in the middle row of
$B_2$ diagrams above, and also for the image of this diagram under
$\inc$ (or arbitrary compositions of $\inc$).  Let $w$ be a sequence
in $\{r,b \}$ of length $n-2m$, then write $e^r_w$ for the following
element of $B_n$:
\[
e^r_w = \raisebox{-25pt}{
\includegraphics{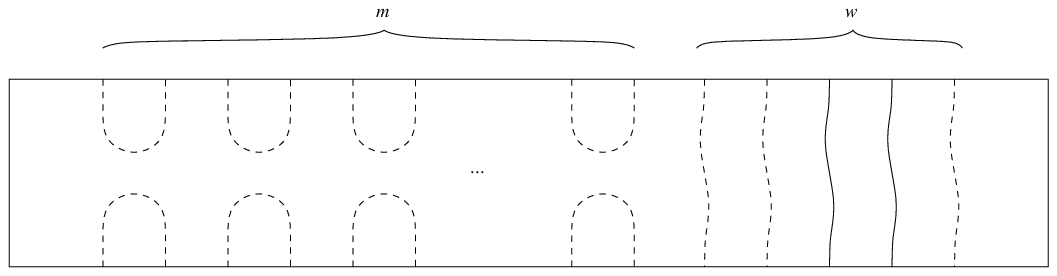}}
\]
(in this example $w=rrbbr$). 

\section{Solutions to the Yang-Baxter Equations}
\setcounter{paragraph}{0}

In the sections {\em after}\/ this one we will return to discussing
the general algebra basis and irreducible representation theory.
First let us briefly look at how the algebra can be used to build
solutions to the Yang-Baxter equations. We will need to start with
some notation and definitions.

\subsection{One colour notations}

Consider the ordinary spin chain representation \cite{TL71,Baxter} of
the one colour Temperley-Lieb algebra.  Choose a basis $v_{1}=|+
+\rangle$, $v_{2} =|+ -\rangle$, $v_{3} =|- +\rangle$, $v_{4} =|-
-\rangle$ and define
\begin{equation}
\label{TLrep}
e = 
\pmatrix{\raisebox{-4pt}{\includegraphics{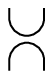}}} =
\pmatrix{
0&0&0&0\cr
0&q&1&0\cr
0&1&q^{-1}&0\cr
0&0&0&0}
\end{equation}
We can think of this as a product of a bra and a ket
\begin{equation}
\label{TLbraket}
e = 
\pmatrix{\raisebox{-4pt}{\includegraphics{rr.eps}}} =
\pmatrix{\raisebox{-4pt}{\includegraphics{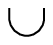}}} 
\pmatrix{\raisebox{-4pt}{\includegraphics{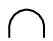}}} = 
\pmatrix{0\cr q^{\frac{1}{2}}\cr q^{-\frac{1}{2}}\cr 0}
\pmatrix{0&q^{\frac{1}{2}}&q^{-\frac{1}{2}}&0}
\end{equation}
where
\begin{equation}
\pmatrix{\raisebox{-4pt}{\includegraphics{er.eps}}}^{t} =  
\pmatrix{\raisebox{-4pt}{\includegraphics{re.eps}}} = 
\pmatrix{0&q^{\frac{1}{2}}&q^{-\frac{1}{2}}&0}
\end{equation}

\subsection{Two colour representation}

We now describe a (highly reducible) representation of the two colour
algebra. This is a representation on the tensor product space
$\mathbb{C}^{4^n}\cong\mathbb{C}^{4}\otimes\mathbb{C}^{4}\otimes\ldots
\otimes\mathbb{C}^{4}$.  We specify the representation by giving the
explicit representation matrices of the ten elements of the basis
$B_{2}$ of $T^{2}_{2}$ on
$\mathbb{C}^{16}\cong\mathbb{C}^{4}\otimes\mathbb{C}^{4}$. The
representation matrices are thus $16\times 16$ matrices, with entries
that now depend on two parameters $q_r$ and $q_b$. The matrices are
given in the basis $v_{1}=|r_+ r_+\rangle$, $v_{2} =|r_+ r_-\rangle$,
$v_{3} =|r_+ b_+\rangle$, $v_{4} =|r_+ b_-\rangle$, $v_{5} =|r_-
r_+\rangle$, $v_{6} =|r_- r_-\rangle$, $v_{7} =|r_- b_+\rangle$,
$v_{8} =|r_- b_-\rangle$, $v_{9} =|b_+ r_+\rangle$, $v_{10}=|b_+
r_-\rangle$, $v_{11}=|b_+ b_+\rangle$, $v_{12}=|b_+ b_-\rangle$,
$v_{13}=|b_- r_+\rangle$, $v_{14}=|b_- r_-\rangle$, $v_{15}=|b_-
b_+\rangle$, $v_{16}=|b_- b_-\rangle$ of $\mathbb{C}^{16}$, where $r$
and $b$ refer to the two colours, and we have an additional variable
living on the lines which you may think of as an arrow pointing up
($+$) or down ($-$), as in the usual spin chain representation of the
\TL\ algebra, compare equation \eref{TLrep}.

The representation matrices for the elements with two propagating straight
red or blue lines are diagonal, with elements
\begin{eqnarray*}
\pmatrix{\raisebox{-4pt}{\includegraphics{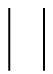}}}&=& 
\mbox{diag}(1,1,0,0,1,1,0,0,0,0,0,0,0,0,0,0)\, ,\\
\pmatrix{\raisebox{-4pt}{\includegraphics{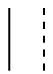}}}&=& 
\mbox{diag}(0,0,1,1,0,0,1,1,0,0,0,0,0,0,0,0)\, ,\\
\pmatrix{\raisebox{-4pt}{\includegraphics{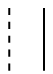}}}&=& 
\mbox{diag}(0,0,0,0,0,0,0,0,1,1,0,0,1,1,0,0)\, ,\\
\pmatrix{\raisebox{-4pt}{\includegraphics{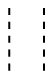}}}&=& 
\mbox{diag}(0,0,0,0,0,0,0,0,0,0,1,1,0,0,1,1)\, ,\\
\end{eqnarray*}
those for two crossing lines of different colour are
\begin{eqnarray*}
\fl
\pmatrix{\raisebox{-4pt}{\includegraphics{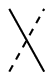}}}=
\pmatrix{\raisebox{-4pt}{\includegraphics{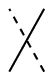}}}^{t}=
=\pmatrix{
0&0&0&0&0&0&0&0&0&0&0&0&0&0&0&0\cr
0&0&0&0&0&0&0&0&0&0&0&0&0&0&0&0\cr
0&0&0&0&0&0&0&0&0&0&0&0&0&0&0&0\cr
0&0&0&0&0&0&0&0&0&0&0&0&0&0&0&0\cr
0&0&0&0&0&0&0&0&0&0&0&0&0&0&0&0\cr
0&0&0&0&0&0&0&0&0&0&0&0&0&0&0&0\cr
0&0&0&0&0&0&0&0&0&0&0&0&0&0&0&0\cr
0&0&0&0&0&0&0&0&0&0&0&0&0&0&0&0\cr
0&0&1&0&0&0&0&0&0&0&0&0&0&0&0&0\cr
0&0&0&0&0&0&1&0&0&0&0&0&0&0&0&0\cr
0&0&0&0&0&0&0&1&0&0&0&0&0&0&0&0\cr
0&0&0&0&0&0&0&0&0&0&0&0&0&0&0&0\cr
0&0&0&1&0&0&0&0&0&0&0&0&0&0&0&0\cr
0&0&0&0&0&0&0&0&0&0&0&0&0&0&0&0\cr
0&0&0&0&0&0&0&0&0&0&0&0&0&0&0&0\cr
0&0&0&0&0&0&0&0&0&0&0&0&0&0&0&0}\, .
\end{eqnarray*}
Finally, there are four colourings of the usual Temperley-Lieb
generators.  These can again be written as products of kets and bras,
just as in equation \eref{TLbraket} for the one colour case,
\begin{eqnarray*}
\pmatrix{\raisebox{-4pt}{\includegraphics{rr.eps}}}&=& 
\pmatrix{\raisebox{-4pt}{\includegraphics{er.eps}}} 
\pmatrix{\raisebox{-4pt}{\includegraphics{re.eps}}}\, ,\\
\pmatrix{\raisebox{-4pt}{\includegraphics{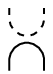}}}&=& 
\pmatrix{\raisebox{-4pt}{\includegraphics{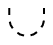}}} 
\pmatrix{\raisebox{-4pt}{\includegraphics{re.eps}}}\, ,\\
\pmatrix{\raisebox{-4pt}{\includegraphics{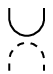}}}&=& 
\pmatrix{\raisebox{-4pt}{\includegraphics{er.eps}}} 
\pmatrix{\raisebox{-4pt}{\includegraphics{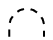}}}\, ,\\
\pmatrix{\raisebox{-4pt}{\includegraphics{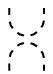}}}&=& 
\pmatrix{\raisebox{-4pt}{\includegraphics{eb.eps}}} 
\pmatrix{\raisebox{-4pt}{\includegraphics{be.eps}}}\, ,
\end{eqnarray*}
where now
\begin{eqnarray*}
\fl
\pmatrix{\raisebox{-4pt}{\includegraphics{er.eps}}}^{t} &=& 
\pmatrix{\raisebox{-4pt}{\includegraphics{re.eps}}} \;=\;
\pmatrix{0&q_{r}^{\frac{1}{2}}&0&0&q_{r}^{-\frac{1}{2}}&0&0&0&0&0&0&0&0&0&0&0}
\, ,
\\
\fl
\pmatrix{\raisebox{-4pt}{\includegraphics{eb.eps}}}^{t} &=& 
\pmatrix{\raisebox{-4pt}{\includegraphics{be.eps}}} \;=\;
\pmatrix{0&0&0&0&0&0&0&0&0&0&0&q_{b}^{\frac{1}{2}}&0&0&
q_{b}^{-\frac{1}{2}}&0}
\, .
\end{eqnarray*}

\subsection{Yang-Baxter construction}

Let us elaborate on this by means of an explicit example.  {}From
certain representations of the two-colour bubble algebra, we can
derive integrable vertex models.  In the simplest scenario, these
vertex models correspond to $\check{R}$-matrix solutions of the
Yang-Baxter equation
\begin{equation}
\label{ybe}
\fl
\left(\check{R}(u)\otimes \mbox{id}\right)
\left(\mbox{id}\otimes\check{R}(u+v)\right)
\left(\check{R}(v)\otimes \mbox{id}\right)
=
\left(\mbox{id}\otimes\check{R}(v)\right)
\left(\check{R}(u+v)\otimes \mbox{id}\right)
\left(\mbox{id}\otimes\check{R}(u)\right)
\end{equation}
which is an equation on a triple tensor product space $V\otimes
V\otimes V$, with $\check{R}$ acting on $V\otimes V$. A particular
vertex model on the square lattice is specified by the matrix elements
of $\check{R}(u)$, which correspond to the Boltzmann weights of the
respective local configurations (the variable $u$ is the spectral
parameter).  The transfer matrices $\mathcal T_{\Nsites}(u)$ are built
from elementary matrices $\check{R}(u)$, so as to act on the
$\Nsites$-fold tensor product $V\otimes V\otimes \ldots \otimes V$.
They commute for different values of $u$.  The free energy of the
vertex model is then obtained from the largest eigenvalue of the
transfer matrix, and we are particularly interested in its behaviour
as $\Nsites$ tends to infinity.

The $\check{R}$-matrix
\begin{equation}
\check{R}(u) = 
\frac{\sin(\lambda-u)}{\sin(\lambda)} I + \frac{\sin(u)}{\sin(\lambda)} e
= 
\frac{\sin(\lambda-u)}{\sin(\lambda)} 
\pmatrix{\raisebox{-4pt}{\includegraphics{rrd.eps}}}
 + \frac{\sin(u)}{\sin(\lambda)} 
\pmatrix{\raisebox{-4pt}{\includegraphics{rr.eps}}},
\end{equation}
expressed in terms of generators of the (one-colour) Temperley-Lieb
algebra with $q+q^{-1}=2\cos{\lambda}$, is a well-known example of a
solution of the Yang-Baxter equation \cite{WDA89}.  In fact it
follows from the relations in the Temperley-Lieb algebra that this
combination satisfies the Yang-Baxter equation.  Hence this
`Baxterisation' shows that any representation of the Temperley-Lieb
algebra with a large $\Nsites$ limit yields a solvable lattice model
of Statistical Mechanics.  For the representation at hand, this model
is the well-known six-vertex model, and the $\check{R}$-matrix is
related to the affine Lie algebra $A_{1}^{(1)}$ \cite{J86}.

Let us now move on to the two-colour case.  The $\check{R}$-matrix
\cite{GP93,GW95b}
\begin{eqnarray}
\fl
\check{R}(u) &=& 
\frac{\sin(\lambda-u)\sin(3\lambda-u)}{\sin(\lambda)\sin(3\lambda)}
\left[
\pmatrix{\raisebox{-4pt}{\includegraphics{rrd.eps}}}+
\pmatrix{\raisebox{-4pt}{\includegraphics{bbd.eps}}}
\right]
+\frac{\sin(3\lambda-u)}{\sin(3\lambda)}
\left[
\pmatrix{\raisebox{-4pt}{\includegraphics{rbd.eps}}}+
\pmatrix{\raisebox{-4pt}{\includegraphics{brd.eps}}}
\right]\nonumber\\
\fl
&&-\frac{\sin(u)\sin(2\lambda-u)}{\sin(\lambda)\sin(3\lambda)}
\left[
\pmatrix{\raisebox{-4pt}{\includegraphics{rr.eps}}}+
\pmatrix{\raisebox{-4pt}{\includegraphics{bb.eps}}}
\right]
+\frac{\sin(u)}{\sin(3\lambda)}
\left[
\pmatrix{\raisebox{-4pt}{\includegraphics{rb.eps}}}+
\pmatrix{\raisebox{-4pt}{\includegraphics{br.eps}}}
\right]
\nonumber\\
\fl
&&+\frac{\sin(u)\sin(3\lambda-u)}{\sin(\lambda)\sin(3\lambda)}
\left[
\pmatrix{\raisebox{-4pt}{\includegraphics{rbx.eps}}}+
\pmatrix{\raisebox{-4pt}{\includegraphics{brx.eps}}}
\right]
\label{baxterisation}
\end{eqnarray}
with $q+q^{-1}=-2\cos(\lambda)$, satisfies the Yang-Baxter equation
\eref{ybe} as a consequence of the relations of the bubble algebra.
Thus any representation of the two-colour bubble algebra on a tensor
product space $V\otimes V\otimes \ldots \otimes V$ with
$q_{r}=q_{b}=:q$ gives rise to an integrable vertex model with an
$\check{R}$ matrix given by equation \eref{baxterisation} with
Boltzmann weights that are trigonometric functions of the spectral
parameter $u$.
 
For the representation at hand, the $\check{R}$ matrix turns out to be
related to the affine Lie algebra C$_{2}^{(1)}$ \cite{J86}. It differs
from the vertex model of \cite{J86} by a
spectral-parameter-dependent gauge transformation \cite{GP93}.

\section{Combinatorics and representation theory}\label{Sbasis}
\setcounter{paragraph}{0}

\subsection{The algebra basis $B_n$}

Note that $B_n$ is somewhat like the Brauer diagram basis $\BBase_n$
of the Brauer algebra $J_n$ \cite{B37}, which in turn contains the
diagram basis of $T_n$.  From each `seed' element of the Brauer basis
we can get some number (0 or more) of diagrams of $B_n$ by colouring
the lines in the following way.  First put a total order on the $n$
lines (any one will do --- for definiteness we will number the line
coming out of the top left endpoint 1, then number other lines
2,3,.. as they first appear reading clockwise round the frame).  Each
line, considered in this order, may be coloured red or blue, unless it
crosses one or more already coloured lines, in which case it must be
coloured in a colour {\em distinct}\/ from those of all the crossing
lines. Of course, if there are three or more lines in the diagram this
may not be possible (i.e.\ when a line crosses both of a pair of
crossed lines), in which case there are {\em no}\/ coloured Brauer
diagrams of this type in $B_n$.

Apart from the seeds with no possible colourings, each seed produces
$2^c$ elements of $B_n$, where $c$ is the number of lines in the seed
which are either without any crossings, or are the first line in their
crossed cluster (i.e.\ the line whose colour is chosen freely).

\paragraph{}
As with $T_n$ \cite{M91} there is a bra-ket construction.  Imagine
cutting the bubble in half (i.e.\ cutting through the front and back
sheets, leaving a top and bottom piece each with Y cross-section).
It will be evident that it is possible to do this such that only
propagating lines are cut, and these once each.  Note that given two
pieces in this way, because of the non-crossing within a layer rule,
there is a unique way of recombining them, i.e.\ recovering the
original diagram.  Indeed any bra- (top piece) and -ket (bottom
piece) such that the number of cut lines matches up (on each sheet of
the bubble separately) may be combined in a unique way.  For example
\[
\includegraphics{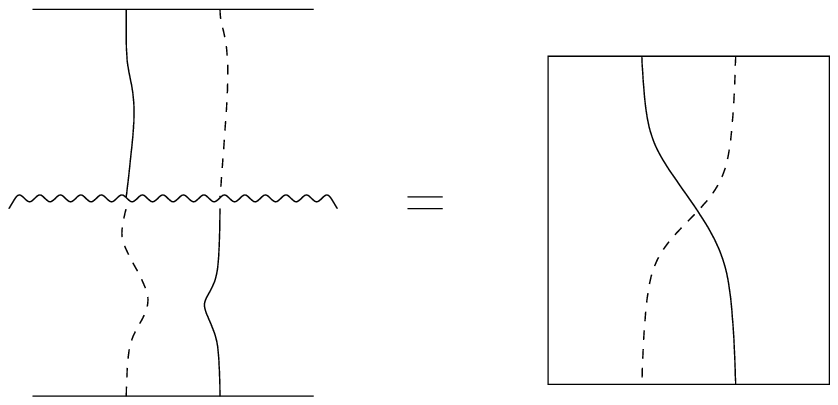}
\]
That is, writing $B_{n}^{\bra{}}(i,j)$ for the set of bra pieces
obtained by cutting elements of $B_{n}(i,j)$ (and similarly
$B_{n}^{\ket{}}(i,j)$ for the set of ket pieces), then any element $a$
of $B_{n}^{\bra{}}(i,j)$ may be combined with any element $b$ of
$B_{n}^{\ket{}}(i,j)$ in a unique way to make a diagram $ab$ in
$B_{n}(i,j)$:
\begin{equation}\label{braxket}
B_{n}(i,j) \cong B_{n}^{\bra{}}(i,j) \times B_{n}^{\ket{}}(i,j) .
\end{equation}

\paragraph{}
Following \cite[\S13.2]{M91}, we define certain injective
homomorphisms of bra sets for any appropriate $n,i,j$:
\[ 
\opA_r : B_{n-1}^{\bra{}}(i,j) \hookrightarrow B_{n}^{\bra{}}(i+1,j)
\]
takes a diagram $d$ to a diagram differing from $d$ only in having an
additional red propagating line at the right hand end;
\[ 
\opA_b : B_{n-1}^{\bra{}}(i,j) \hookrightarrow B_{n}^{\bra{}}(i,j+1)
\]
similarly, but adding a blue line; for $i>0 $ 
\[ 
\opB_r : B_{n-1}^{\bra{}}(i,j) \hookrightarrow B_{n}^{\bra{}}(i-1,j)
\]
takes a diagram $d$ to a diagram differing from $d$ only in having the
Southern endpoint of the last (rightmost) red propagating line in $d$
turn back to form a new rightmost Northern vertex; and $\opB_b$
similarly for the last blue line.

It will be evident that 
\begin{equation}\label{preres}
\fl
B_{n}^{\bra{}}(i,j) = 
\opA_r B_{n}^{\bra{}}(i-1,j) \cup \opA_b B_{n}^{\bra{}}(i,j-1) 
\cup \opB_r B_{n}^{\bra{}}(i+1,j)
\cup \opB_b B_{n}^{\bra{}}(i,j+1)
\end{equation}
(any undefined set here to be interpreted as the empty set). 

\paragraph{}
It will be convenient to be able to depict the sum of two diagrams in
$B_n$ differing only in the colour of one line by drawing any one of
these diagrams with the relevant line replaced by a thick line (called
a {\em white}\/ line). We will generalise this so that a diagram with
two white lines is a sum of four diagrams from $B_n$, and so on.  Let
us write $U_i$ for that diagram which has the same shape as the
diagram $U_{i} \in T_n$, but has all white lines.  Note that this is
an (unnormalised) idempotent: $U_i U_i = (\delta_r + \delta_b) U_i$.

Write $e_l$ for an all white diagram of the same shape as $e^r_w$,
where $l$ is the length of the sequence $w$. Thus $e_{n-2m} = U_1 U_3
.. U_{2m-1}$.

Let $\tilde{n}$ denote the element of $\{ 0,1 \}$ congruent to $n$
modulo 2.

\subsection{Standard modules}

Next we construct a basic set of representations of $T^2_n$. 

\paragraph{}
It will be evident from equation \eref{prop_no} that there is a
filtration of $T^2_n$ by ideals with sections spanned by diagrams
having fixed numbers of propagating lines --- and hence having the
subsets $B_n(i) \subset B_n$ as bases.

Note in particular from equation \eref{prop_no} that $T^2_n U_1 T^2_n$
has basis $B_n[n-2] = \cup_{i \leq n-2} B_n(i)$; $T^2_n U_1 U_3 T^2_n$
has basis $B_n[n-4]$; and so on.  (If $n$ is big enough then $T^2_n
U_1 T^2_n = T^2_n U_1 U_2 T^2_n $, $T^2_n U_1 U_3 T^2_n = T^2_n U_1
U_3 U_2 U_4 T^2_n $, and so on, so these ideals may be considered
idempotently generated over any field $K$.)  The filtration by
propagating lines may thus be written
\[
T^2_n \supset T^2_n U_1 T^2_n \supset T^2_n U_1 U_3 T^2_n 
\supset .. \supset T^2_n U_1 U_3 .. U_{2m-1} T^2_n
\supset ..
\]
Let us write $T^2_n[i]$ for the ideal spanned by diagrams with $\leq
i$ propagating lines.

The total number $i+j$ of propagating lines is one of
$n,n-2,n-4,..,1/0$.  This number may be partitioned in any way between
red and blue lines, with the corresponding ideal breaking up as a
direct sum accordingly.  The filtration thus refines to one with
sections spanned by the sets $B_n(i,j)$ of diagrams with fixed {\em
propagating index}\/ $(\hash_{r}(d),\hash_{b}(d))$.

Each such section breaks up as a sum of isomorphic left modules each
with basis of the form $\{ \bra{a}\ket{b} \;\; | \;\; \bra{a} \in
B_{n}^{\bra{}}(i,j) \}$ where $\bra{a}$ varies over all possibilities
and $b$ is fixed.  (There is obviously a parallel construction for
right modules.)  We denote (any representative of) the equivalence
class of these summands $\Delta_n(i,j)$.  These left modules are
called {\em standard}\/ modules.

For example, $\Delta_2(1,1)$ has basis
\[
\includegraphics{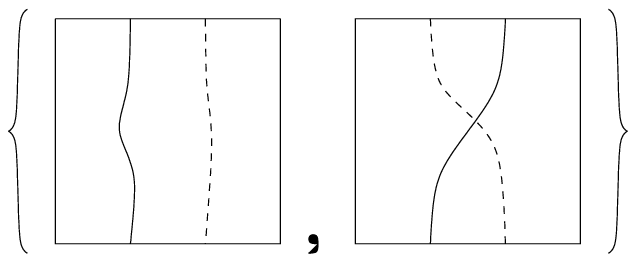}
\]
while $\Delta_2(2,0)$ has basis
\[
\includegraphics{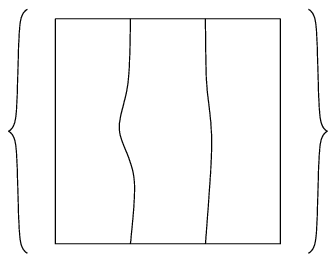}
\]
Note that, because of the sectioning, the action of $B_2(0,0)$
elements on this object {\em regarded as a basis element of}\/
$\Delta_2(2,0)$ is to give {\em zero}\/ (that is, any object with
fewer propagating lines would lie in the next layer of the filtration,
and is thus congruent to zero in this section).

\paragraph{}
Since the bottom (ket) halves of diagrams regarded as basis elements
of standard modules play no role, we have another basis for each
standard module $\Delta_{n}(i,j)$, consisting of the set
$B_{n}^{\bra{}}(i,j)$ of bra diagrams, with the action defined in an
obvious way.

Note the following. 

\paragraph{}
The construction of standard modules is independent of the choice of
field.

Each standard module $\Delta_{n}(i,j)$ comes with an inner product via
its basis of bra diagrams (and dual basis of ket diagrams):
\begin{equation}\label{dodah}
\begin{array}{lllll}
d & d' & = & k_{dd'} & d'' 
\\ 
\bra{d}\ket{d} & \bra{d'} \ket{d'} & = &  \ket{d}\bra{d'} & \bra{d}\ket{d'}
\end{array}
\end{equation}
In particular, if $i+j=n$, it is easily verified that the
corresponding Gram matrix, $G_n(i,j)$, is the unit matrix. Thus
$\Delta_{n}(i,n-i)$ is irreducible for any $q_r, q_b$.

Note on the other hand that $G_2(0,0) = \mbox{diag}(\delta_r ,
\delta_b)$, so that $\Delta_2(0,0)$ is reducible if either $ \delta_r$
or $\delta_b$ vanishes.

More generally, it will be evident that $|G_n(i,j)|$ is a non-zero
polynomial in $\delta_r, \delta_b$. Thus
\begin{theo}\label{theo1}
The standard modules $\Delta_n(i,j)$ are generically 
simple.
\end{theo}
(Recall that {\em generically}\/ means: in a Zariski open subset of
the $(\delta_r,\delta_b)$ parameter space.)

On the other hand, inspection of the diagram for $e^r_w$ shows the
following:
\begin{pr}\label{(stngen)}
Let $\delta_r$ be invertible in $K$ and let $e^r_w \in T^2_n$ have
sequence $w=rr..rbb..b=r^ib^j$. Then
\[
\Delta_n(i,j) \cong T^2_n e^r_w \hspace{1in} \mbox{mod.\ $T^2_n[i+j-2]$}
\]
\end{pr}
\begin{pr}\label{}
Over any field $K$ with $\delta_r$ invertible $\Delta_n(i,j)$ has
simple head.
\end{pr}
(Recall that the {\em head}\/ of a module is the quotient by the
intersection of all maximal proper submodules.)

\noindent{\em Proof:} The generating element provided by the previous
proposition is an unnormalised (but normalisable) idempotent. It is
not primitive in $T^2_n$, but it is primitive in a suitable
quotient. This means that the induced module is indecomposable
projective in some quotient, so it has a simple head. Done.

Next we consider the {\em completeness}\/ of this set of
representations.

\paragraph{}
Irrespective of the choice of field, the restriction of standard
module $\Delta_n(i,j)$ to $T^2_{n-1}$ works as follows.  If the line
coming out of the last (rightmost) northern endpoint is propagating
and red (resp.\ blue), then $d$ behaves, on restriction, like an
element of the corresponding basis of $\Delta_{n-1}(i-1,j)$ (resp.\
$\Delta_{n-1}(i,j-1)$).  That is, these modules are submodules of
$\Res{n-1}{n}{\Delta_{n}(i,j)}$.  If the line coming out of the last
(rightmost) northern endpoint is red (resp.\ blue) and not
propagating, then, {\em quotienting by the submodules just noted}, $d$
behaves, on restriction, like an element of the corresponding basis of
$\Delta_{n-1}(i+1,j)$ (resp.\ $\Delta_{n-1}(i,j+1)$).  For example in
$\Delta_{4}(2,0)$, restricting to $n=3$ as indicated by the brace we
have:
\[
\includegraphics{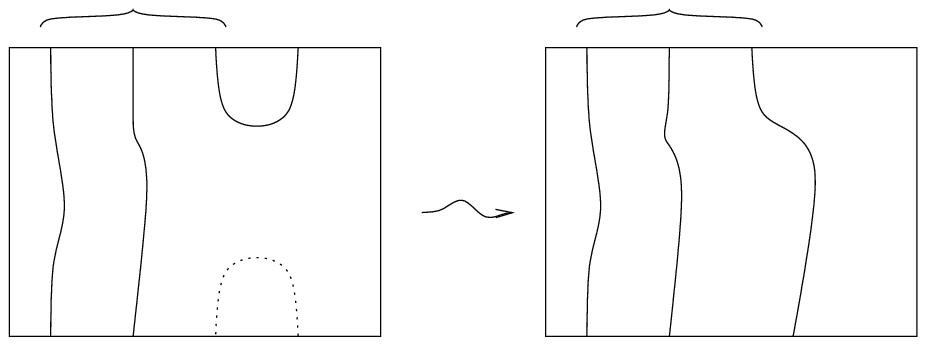}
\]

This information is neatly summarised (including positivity
constraints) as follows. Define a bipartite infinite graph $\bigra$
with vertex set $\N_0 \times \N_0$ by
\[
\includegraphics{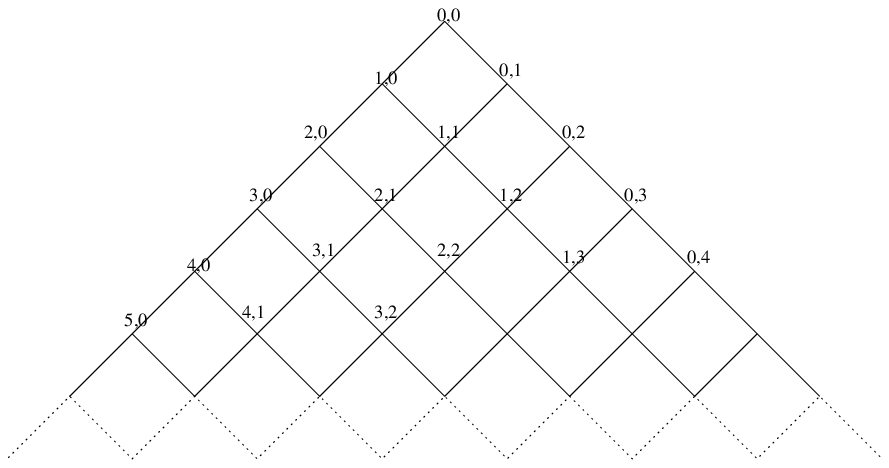}
\]
Define $\bigra_n$ to be the full subgraph with vertices $(i,j)$ such
that $i+j \leq n$.
\begin{pr}\label{res1}
The restriction from $T^2_n$ to $T^2_{n-1}$ of $\Delta_{n}(\mu)$ may
be given by   
\[
\Res{n-1}{n}{\Delta_{n}(\mu)} 
\cong  \bigplus_{\lambda } \; \Delta_{n-1}(\lambda)
\]
where the sum is over the set of 
nearest neighbours of $\mu=(i,j)$ on the graph $\bigra_{n}$. 
\end{pr}

It follows that the dimension of the module $\Delta_n(i,j)$ is the
number of walks from $(0,0)$ to $(i,j)$ of length $n$.  It follows
from this and equation \eref{braxket} that the rank of the algebra
(the degree of $B_n$) is the sum of the squares of these dimensions;
and that
\begin{theo}
If the algebra is semisimple in some specialisation of the parameters
(as generically, for example --- see Theorem~$\ref{theo1}$), then the
standard modules are a complete set of irreducible modules in that
specialisation.
\end{theo}

\paragraph{}
A finite dimensional algebra (call it $A$) has, of course, finitely
many classes of irreducible representations.  Let $\Simp(A)$ be the
set of these.  Now suppose $B$ is a subalgebra of $A$, with
irreducibles $\Simp(B)$.  We can restrict $R \in \Simp(A)$ to be a
representation of $B$ (every element of $B$ has a representation
matrix in $R$ since $B \subset A$).  As a representation of $B$ this
$R$ will not in general be irreducible --- in general it will be made
up of a sum of one or more $B$ irreducibles (we say, one or more
factors).  A {\em Bratteli diagram}\/ of the pair $(A,B)$ is a graph
with a vertex for each element of $\Simp(A)$ and a vertex for each
element of $\Simp(B)$ (the union of these vertices is usually but not
always taken disjoint).  If there is an edge between $a$ and $b$ say,
then it means the restriction of $a \in \Simp(A)$ contains $b$ as a
factor.  Thus in particular the fact that ``the sum of the dimensions
of all the factors of $a$ is the dimension of $a$ itself'' becomes
``the dimension at $a$ is the sum of the dimensions of the nearest
neighbours of $a$ on the graph (possibly with multiplicities)''.

A Bratteli diagram of a tower of algebras $A \supset B \supset C ...$
extends this in the obvious way.  If the last algebra in this sequence
is a one-dimensional algebra (with one 1d irrep, call it $o$) then
the number of walks on the graph from $o$ to point $a$ will be the
dimension of the irreducible $a$ (so the set of these walks will be a
basis for $a$).  Again for the tower, $\Simp(A)$ and $\Simp(C)$ may
(for ease of drawing, say) be allocated some vertices in common
(`foreshortened' diagram).  This does not spoil the counting if we are
careful (each vertex corresponds to irreducibles in more than one
algebra, but, on fixing a given algebra, that vertex becomes
unambiguous).

\paragraph{}
In our algebra we see that the standard modules play a special role,
somewhat akin to that of simple modules, although over an arbitrary
field they are not themselves necessarily simple.

The Bratteli {\em basis diagram}\/ of such an algebra is a certain
embellished graph, each vertex of which corresponds to a standard
module label (in this case the label consists of $n$ and a propagating
index $\lambda$); and each edge of which corresponds to a factor in
the restriction of that module to $n-1$ (as above).  Each vertex is
embellished with a depiction of a basis for the corresponding standard
module.

A {\em foreshortened}\/ Bratteli basis diagram is a Bratteli basis
diagram viewed from such a direction as to cause certain vertices to
coincide.  In our case it is those vertices for different $n$ which
have the same propagating index (thus a whole tower of embellishments
will have to be drawn at the same point, but there are good reasons
for this --- see later).  Such a diagram, where {\em possible}\/ to
draw, contains essentially complete information on the generic
representation theory of the algebra.  The (foreshortened) Bratteli
basis diagram for $T^2_-$ begins
\begin{equation}\label{fsbrat}
\fl\includegraphics{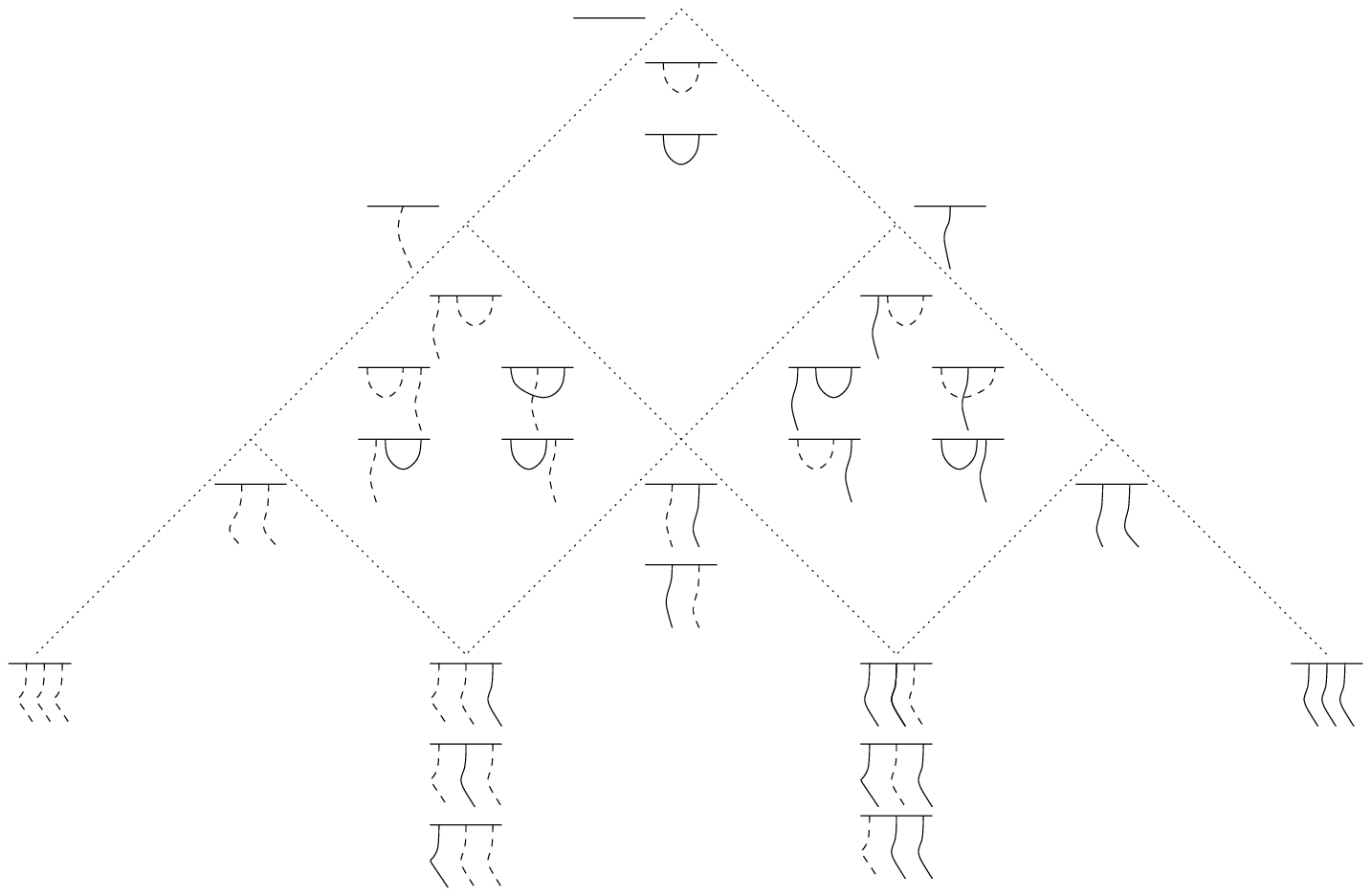}\hspace{-8pt}
\end{equation}
It may be helpful to emphasise that here the basis element
\[
\includegraphics{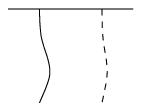}
\]
could also be written
\[
\includegraphics{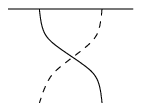}
\]
--- in the bra form these are equivalent, since the red and blue lines
    are not on the same bubble layer.

If an algebra is semisimple (as ours is generically) then its total
dimension is the sum of the squares of the dimensions of the
irreducibles.  Here we see, for any $K$, that the total dimension is
the sum of the squares of the dimensions of the standards.  Thus the
entire combinatoric is encoded in the pictures.  Every possible
diagram is built bra-ket from the kets in the foreshortened Bratteli
diagram (and their descendents).

\paragraph{}
In the ordinary \TL\ case it is the non-semisimple exceptions ($q$
root of unity) which are of most interest. We conclude by setting up
machinery to investigate this case (again paralleling Martin's usual
approach \cite{M96,MW00} to \TL\ and its generalisations).

\section{Categories, roots of unity, conformal series etc.}
\setcounter{paragraph}{0}

We may use a little category theory to very efficiently rederive the
generic representation theory of $T^2_n$ given above, in such a way
that it can readily be extended to the exceptional cases.

Recall $U_i$ from section~\ref{Sbasis} and let $U=U_{n-1}$.  Note that
\begin{equation}\label{main}
U T^2_n U \cong T^2_{n-2}  .  
\end{equation} 
A diagrammatic version of this follows from the representation of $U$
by
\[
 \includegraphics{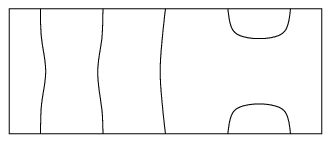}
\]
(all lines `white') so that 
\[
\includegraphics{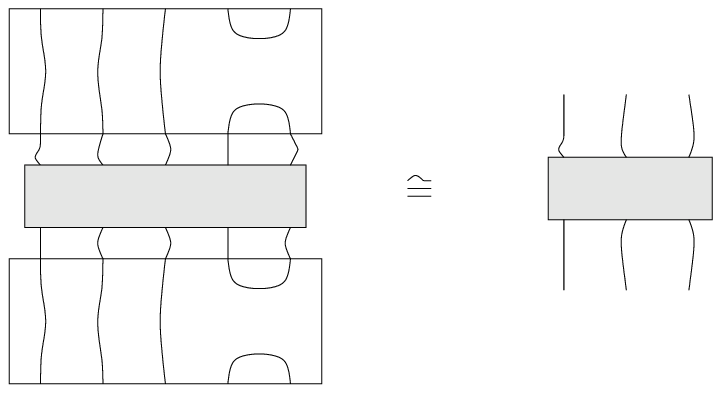}
\]

Note that $U$ commutes with $T^2_{n-2} \subset T^2_{n}$ so that $T^2_n
U$ is a left $T^2_n$ right $T^2_{n-2}$-bimodule.  Let
$F:T^2_{n}\mbox{-mod} \rightarrow T^2_{n-2}\mbox{-mod}$ be the
functor
\[ F:M \rightarrow UM
\]
and $G:T^2_{n-2}\mbox{-mod} \rightarrow T^2_{n}\mbox{-mod}$
\[ G:N \rightarrow T^2_n U \otimes_{T^2_{n-2}} N\, .
\]
It follows from \eref{main} that $FG=1_{T^2_{n-2}\mbox{-mod}}$.  Thus
(so long as $U$ may be normalised as an idempotent) we have a full
embedding of the category $T^2_{n-2}\mbox{-mod}$ in
$T^2_{n}\mbox{-mod}$.  The simple modules $L$ in
$T^2_{n}\mbox{-mod}$ {\em not}\/ hit in this embedding are those for
which $UL=0$.  That is, they are also the simple modules of the
quotient algebra $T^2_n / T^2_n U T^2_n$.

To reiterate, equation \eref{main} gives what is called a full
embedding of $T^2_{n-2}$ in $T^2_n$.  This means that there is a
natural injection of $\Simp(T^2_{n-2})$ into $\Simp(T^2_n)$ ---
another reason for the foreshortening of the Bratteli diagram.  It
also means that most of the representation theory of $T^2_n$ follows
from that of $T^2_{n-2}$ via a little elementary category theory ---
and hence inductively from the trivial cases $T^2_0$ and $T^2_1$.

Note that $ T^2_n U T^2_n$ includes every diagram except those with
exactly $n$ propagating lines.  Thus $T^2_n / T^2_n U T^2_n$ is
spanned by the set $B_n(n)$ of diagrams with exactly $n$ propagating
lines.  Let $\Gamma_n$ denote an index set for the simple modules of
$T^2_n$, and $\Lambda_n$ an index set for the quotient algebra $T^2_n
/ T^2_n U T^2_n$.  So long as $U$ may be renormalised as an idempotent
it follows that
\begin{equation}\label{dag1}
\Gamma_n = \Gamma_{n-2} \cup \Lambda_n
\end{equation}
where the union is disjoint.  The full embedding allows us to inject
$\Gamma_{n-2} \hookrightarrow \Gamma_n$ in precisely the way implied
by the foreshortening of our foreshortened Bratteli diagram.

By equation \eref{dag1}, we know $\Gamma_n$ if we know $\Lambda_m$ for
all $m$. We now determine this set.

\subsection{The subalgebra generated by $B_n(n)$}
\label{funky}

Note that $KB_n(n)$ is a subalgebra of $T^2_n$.  Let $\sT_n$ denote
this subalgebra, then
\[
\sT_n \hookrightarrow T^2_n \doublerightarrow T^2_n / T^2_n U T^2_n
\]
is a sequence of algebra morphisms. The composite morphism takes an
element of $B_n(n)$ to the same object regarded as a basis element of
$T^2_n / T^2_n U T^2_n$, hence it is an isomorphism.  Provided our
ground field $K$ has characteristic different from two (we are mainly
interested in $\C$, with characteristic zero, of course) this algebra
may be identified with a certain quotient of the wreath product group
algebra $K C_2 \wr S_n$, where $C_{2}$ denotes the cyclic group of
order two and $S_{n}$ the symmetric group (permutation group of $n$
elements) of order $n!$, as follows.

Elements of $C_2 \wr S_n$ may be represented in the form of
permutation diagrams where each line carries zero or one beads:
\[
\includegraphics{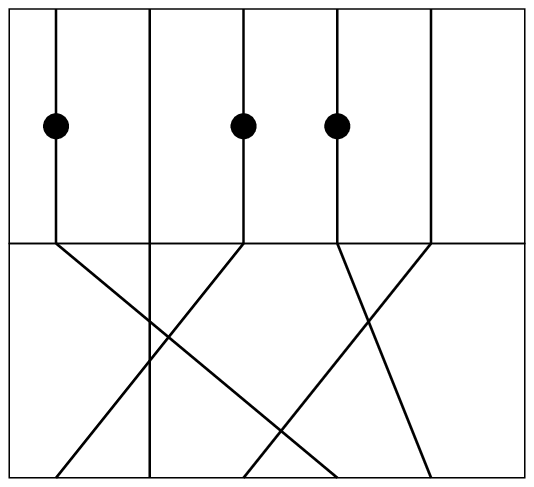}
\]
(i.e.\ there are $2^n n!$ such diagrams).  The rule of composition is
then as for ordinary permutations except that the number of beads on a
single line is reduced modulo 2.

Such a diagram $d$ in which some line is replaced by a (beadless) red
(resp.\ blue) line is to be understood as the linear combination
\begin{equation}\label{d'}
 d' = \frac{d_0 \pm d_1}{2} 
\end{equation}
where $d_0,d_1$ denote the diagram with that line having zero/one
beads respectively.  Replacing all lines in this way, in all possible
ways, we have another basis of $K C_2 \wr S_n$ consisting of all
possible two-colourings of permutations.  The composition rule here
(in consequence of \eref{d'}) is to compose permutations by
juxtaposition as usual, except that if two different coloured lines
are juxtaposed the composite is zero.

Consider the subset of this basis consisting of elements in which two
lines may cross only if their colour is distinct.  This is a basis for
a subalgebra (to see this again consider the different coloured lines
as living in two different layers --- within a layer there are no
crossings, and this is not affected by composition). Indeed it will be
evident that this subalgebra may be {\em identified}\/ with $\sT_n$.

Now define two linear combinations in $KC_2 \wr S_2$, each of shape
\[
\includegraphics{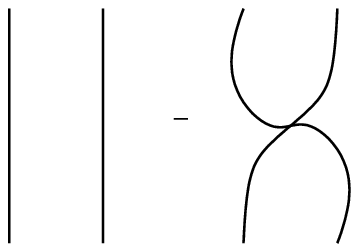}
\]
but one with all lines coloured red, one with all lines coloured blue.
Note that these are (unnormalised) idempotents.  Define algebra
$H^2_n$ to be the quotient of $KC_2\wr S_n$ (any $n>1$) by both these
objects.

It is well known that the irreducible representations of $KC_2\wr S_n$
over $K=\C$ are indexed by pairs of integer partitions of combined
degree $n$, and that the idempotents above are the primitive and
central idempotents of $KC_2\wr S_2$ corresponding to the
one-dimensional irreducible representations indexed by $((1^2),)$ and
$(,(1^2))$.  It follows (after a little work, see \cite{MW02}) that
the index set for irreducibles of the quotient $H^2_n$ is the subset
of the set of pairs of integer partitions in which no partition has a
second row. It follows similarly that the Bratteli diagram for the
tower of these algebras as $n$ varies is the Pascal triangle, i.e.\
the simples for given $n$ lie in the $n^{\rm th}$ layer of the Pascal
triangle.

Note that $B_n(n)$ may be regarded as a basis for $H^2_n$, since in
$H^2_n$ any two lines of the same colour which are crossed may be
replaced by the same two lines uncrossed (i.e.\ a local implementation
of the quotient by the diagram above in that colour).  Now consider
the sequence of algebra morphisms
\[
\sT_n \hookrightarrow KC_2\wr S_n \doublerightarrow H^2_n
\]
The image of a basis element in $B_n(n)$ under the composite map is
the same element regarded as a basis element of $H^2_n$. Thus the
composite is an isomorphism.

We have established a sequence of isomorphisms which allows us to
identify the index set for simple modules of $H^2_n$ with
$\Lambda_n$. This reproduces the layer of the Pascal triangle in our
original foreshortened Bratteli diagram and, taken layer by layer,
using equation \eref{dag1} reproduces the whole foreshortened Bratteli
diagram.

\subsection{On the exceptional structure of $T^2_n$}

The $rb$-sequence of a diagram is the sequence of colours of strings,
read off clockwise from the top left hand corner of the frame.  The
standard basis $B_n(\mu)$ may be partitioned into subsets of elements
with the same $rb$-sequence, called $rb$-parts.  Since colours are
orthogonal, the inner product is zero on any pair from $B_n(\mu)$
unless they lie in the same $rb$-part.  Thus the determinant of the
Gram matrix is a product of corresponding block determinants.

It will be evident that the block {\em determinant}\/ depends only on
the number of $r$'s and $b$'s, not their order in sequence.  It is
thus straightforward to determine the roots of the Gram determinants
(which, we recall, are polynomials in $\delta_r,\delta_b$), as
follows.

For $B_{n}(i,j)$ with $i+j=n$ we have $|G_n(i,j)| =1$.  (Recall that a
module is simple unless its Gram matrix is singular, thus all these
modules are simple.)  For $B_{n}(i,j)$ with $i+j=n-2$ all the lines
but one are propagating, so {\em all}\/ lines of one colour (red or
blue) are propagating.  For example, the $rrrrbb$ part of $B_6(2,2)$
has Gram block
\[
\fl\includegraphics{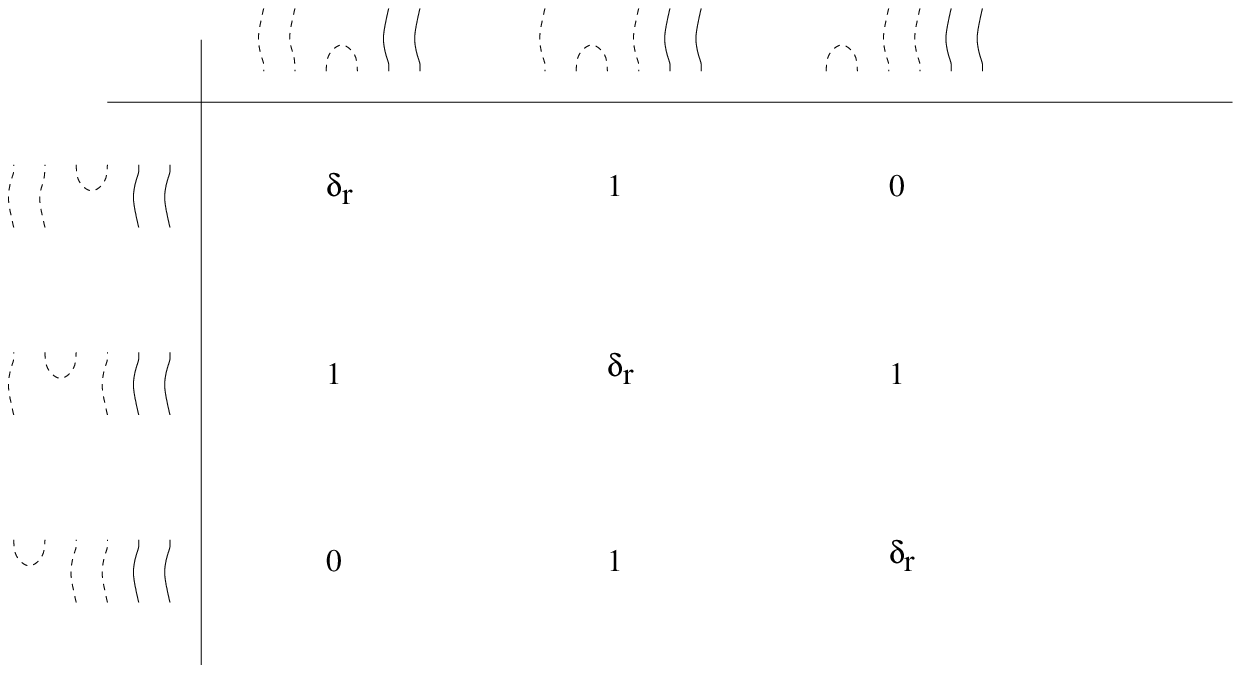}
\]
We see that the colour with all lines propagating plays no role, and
that the Gram block coincides, in this example, with the $(3,1)$ Gram
matrix of ordinary $T_4(\delta_r)$.  Thus the set of roots of {\em
any}\/ such Gram determinant must be taken from the roots of the
\Gram\ determinants of $T_{n}(\delta_r)$ and $T_{n}(\delta_b)$.  These
are well known \cite{M91} to lie in the set of roots of unity (when
expressed in terms of $q_r,q_b$) for each colour.  That is,
\begin{pr}\label{ex1} 
If neither $q_r$ nor $q_b$ is a root of 1 then every module
$\Delta_n(i,j)$ with $i+j=n-2$ is simple.
\end{pr}
\begin{pr}\label{ex3} 
If either $q_r$ or $q_b$ is a root of 1 then there is an $n$ such that
$T^2_n$ is not semisimple.
\end{pr}

Now suppose that for some choice of $K$ some standard module,
$\Delta_n(\mu)$ say, is not simple (as already noted, this has to
happen for the algebra to fail to be semisimple). Then in particular
this module has some simple module, $L({\lambda})$ say, in its socle.
Take $n$ to be at its lowest value such that this occurs.

It is easy to see that both the localisation functor $F$ and the
globalisation functor $G$ take a standard module to a standard module
with the same label (or 0 if no such module exists, in case of $F$).
It is also easy to see that every simple module occurs as the head of
some standard module.  Thus the nonsimplicity of our standard
$\Delta_n(\mu)$ must show up as a morphism of standard modules between
that having the simple $L(\lambda)$ as its head (we might as well call
it $\Delta_n(\lambda)$), and $\Delta_n(\mu)$.  If neither $\mu$ nor
$\lambda$ has $i+j=n$ then we can localise until one of them does,
whereupon the corresponding standard is simple by our earlier
analysis, thus this simple standard must be $\Delta_n(\lambda)$, and
$\Delta_n(\mu)$ must have $i+j <n$, that is to say, $\mu_1 + \mu_2 <
n$.  But now suppose this $\mu_1 + \mu_2 < n-2$ (and there is not a
suitable choice of $\mu$ with $\mu_1 + \mu_2 = n-2$).  Consider the
following {\em Frobenius reciprocity}:
\[
\fl
\Hom(\Ind{n-1}{n}{} \Delta_{n-1}(\lambda_1,\lambda_2 -1)  ,  
\Delta_n(\mu) ) 
\cong 
\Hom(\Delta_{n-1}(\lambda_1,\lambda_2 -1)  ,  \Res{n-1}{n}{} 
\Delta_n(\mu) )
\]
It is straightforward to show that $\Delta_n(\lambda)$ appears in the
head of the induced module $\Ind{n-1}{n}{}
\Delta_{n-1}(\lambda_1,\lambda_2 -1) $.  Thus the left hand hom space
is not empty. But then neither is the right, and we have a nontrivial
homomorphism of distinct standard modules at level $n-1$ also.  This
is a {\em contradiction}\/ of our construction that $n$ is the lowest
value for which such a homomorphism occurs.  Thus we {\em may not}\/
suppose that $\mu_1 + \mu_2 < n-2$, i.e., we must take $\mu_1 + \mu_2
= n-2$.  In other words, the first occurrence of such a morphism must
be into a standard module with this type of label.  But by
proposition~\ref{ex1} such a morphism is only possible if at least one
of $q_r,q_b$ is a root of unity. We have established
\begin{pr}\index{ex2} 
If neither $q_r$ nor $q_b$ is a root of 1, then every module
$\Delta_n(i,j)$ is simple, and $T^2_n$ is semisimple.
\end{pr}

The determination of the complete structure of $T^2_n$ when the
parameters are roots of unity remains for now an open problem.  It
should be amenable to the methods we have developed, but it seems
possible that considerably more donkey work remains.  Given the
connection between the ordinary case and conformal representation
theory (cf. \cite{KooSal,Henkel,MW00}) the answer should raise some
interesting issues.

\section{Conclusions and discussion}
\setcounter{paragraph}{0}

We have constructed the generic irreducible modules of $T^2_n$.  Every
other module (such as occurs in transfer matrices) can be built as a
sum of these.\footnote{ Or in non-semisimple cases as a not
necessarily direct sum of their simple heads.}  Thus the spectrum of
any transfer matrix will be (up to multiplicities) the union of the
spectra computed using these smallest possible modules.  This analysis
thus provides the most efficient tools for explicit computation
(modulo any overarching constraints imposed in practice by, for
example, implementation of the Bethe ansatz),
cf. \cite[\S12.4]{Baxter}.  Note that it also tells us a convenient
labelling scheme for types of correlation functions.  The pair label
$(i,j)$ here replaces the charge sector label relevant for Bethe
ansatz in ordinary spin-chains \cite[\S8.4]{Baxter}.  That is, we
have two naive pseudoparticle types.

In case the second colour is merely to be regarded as a dilution
(i.e.\ it just acts as a placeholder), direct contributions to
correlations involving this colour would be trivial or ignored.  In
more general settings we have the possibility of correlations in which
distinct operators cross over in the plane (a feature which cannot
occur in models built from the ordinary Temperley-Lieb algebra, such
as ordinary Potts models and Ising models).  This `thickening' of the
underlying plane lattice provides scope for considering a number of
further generalisations to more exotic 2d models and possibly also to
3d. We will return to these possibilities in a subsequent paper.

Certainly $T^2_n$ has a number of relatively obvious generalisations
($T^N_n$, $N=3,4,..$, and certain generalisations to more exotic
underlying spaces) for which the corresponding generalised analysis
goes through directly.  Work is in progress to find generalisations of
Grimm and Pearce's original idea accordingly.

It is interesting that the exceptional structure of $T^2_n$ is tied to
the same special parameter choices as are already widely familiar for
2d systems --- $q$ a root of unity --- even though our models have
multiple parameters.  This contrasts sharply with direct attempts to
generalise to 3d, such as the partition algebra, for which a
completely different set of exceptional cases occur \cite{M96}.

Finally we note two points of interest in representation theory.  The
new algebras have features reminiscent of a recent conjecture (see
\cite{MW02}, cf. our section~\ref{funky}) for a basis of generalised
blob algebras.  These algebras have been used recently to probe the
physically relevant part of the representation theory of affine Hecke
algebras \cite{DM02}, so the connection here is intriguing.  Secondly,
we observe that there is no known diagram calculus for the Hecke
algebra quotient associated to $U_qsl_3 $ (in the sense that the
Temperley-Lieb algebra is a Hecke algebra quotient associated to
$U_qsl_2$).  Indeed there is no calculus for $sl_N$ for any $N$ but 2.
Such a calculus has long been sought, and would be enormously useful
in a number of areas of representation theory.  As a generalisation of
the $sl_2$ case, our calculus provides some intriguing clues for
$sl_3$ (although it is {\em not}\/ itself an $sl_3$ calculus).

\ack 

UG would like to express his gratitude to the Erwin Schr\"{o}dinger
International Institute for Mathematical Physics in Vienna for support
during two stays in winter 2002/2003, where part of this work was
done.

\end{document}